\documentclass[aps,prb,twocolumn,floatfix,superscriptaddress,showpacs,10pt]
{revtex4-1}
\usepackage{graphicx}
\usepackage{amssymb}
\usepackage{amsmath}
\makeindex

\newcommand\rs[1]{{\scriptscriptstyle\rm #1}}

\begin{document}

\title{Light-superconducting interference devices}

\author{Frans Godschalk}

\author{Yuli V. Nazarov}
\affiliation{Kavli Institute of Nanoscience, Delft University of Technology,
P.O.  Box 5046, 2600 GA Delft, The Netherlands}

\date{2013}

\begin{abstract}
Recently, we have proposed the half-Josephson laser (HJL): a device that
combines lasing with superconducting leads, providing a locking between the
optical phase and the superconducting phase difference between the leads. In
this work, we propose and investigate two setups derived from a superconducting
quantum interference device (SQUID), where two conventional Josephson junctions
are replaced by two HJLs. 
In the first setup, the HJLs share the same resonant mode, while in
the second setup two separate resonant modes of the two lasers are coupled
optically. We dub the setup `light-superconducting interference device' (LSID). 
 
In both setups, we find the operating regimes similar to those of a single HJL.
Importantly, the steady lasing field is significantly affected by the magnetic
flux penetrating the SQUID loop, with respect to both amplitude and phase. This
provides opportunities to tune or even quench the lasing by varying a small
magnetic field.

For the second setup, we find a parameter range where the evolution equation for
the laser fields supports periodic cycles. The fields are thus modulated with
the frequency of the cycle resulting in an emission spectrum consisting of a set
of discrete modes. From this spectrum, two modes dominate in the limit of strong
optical coupling. Therefore, the LSID can be also used to generate such
modulated light.
\end{abstract}


\maketitle

\section{Introduction}
In the past decade there has been a rapidly growing interest in devices that
combine semiconductor nanostructures with superconductors. The advantage of this
combination lies in the ability of the current day semiconductor technology to
engineer all kinds of devices and nanostructures. When incorporated in a
Josephson junction, these determine the transport
properties of the junction\cite{DeFranceschi}, which, for example, allow to
manipulate supercurrents\cite{Doh}, realize Majorana states\cite{mourik},
facilitate superradiant emission of light\cite{Asano} or become useful for the
purposes of quantum manipulation\cite{Khoshnegar}.

Recently, we have proposed the so-called half-Josephson laser
(HJL)\cite{godschalk}. It consists of a single quantum emitter with two
superconducting leads biased at voltage $V$, and an optical cavity with resonant
mode at frequency $\approx eV/\hbar$. It emits coherent laser light at a
frequency that is precisely $eV/\hbar$, a half of the Josephson frequency, the
optical phase of this light being locked with the superconducting phase
difference between the leads. A HJL can be viewed as a voltage-biased Josephson
junction. Later, after the HJL proposal, we have investigated HJLs with
multiple emitters, that provide exponentially long coherence times for the
emitted light\cite{godschalk2}, and proposed schemes to reduce noise in the
superconducting phase using optical feedback\cite{godschalk3}.  
   
In this Article, we report a study of a HJL application that is build on one of
the archetypical devices made of Josephson junctions: the superconducting
quantum interference device (SQUID). We consider the {\it d.c.} SQUID
setup\cite{dcsquid}, which is a circuit of two parallel Josephson junctions that
supports a supercurrent up to a certain critical value. The whole structure is a
superconducting loop, which can be threaded by a magnetic flux. 
The presence of flux  makes the phase drops at the Josephson junctions unequal.
As a result, the critical current of the device depends periodically on the
flux\cite{Tinkhambook}, so that the SQUID can be viewed as a flux tunable
Josephson junction. 

The subjects of our study are two SQUID setups where the Josephson junctions are
replaced with HJLs. In the simplest case of optically uncoupled HLJ's, the
effect of the magnetic flux on the superconducting phases, combined with the
phase lock of these to the optical phases in the HJLs, will
lead\cite{recher:10} to an optical analogue of the Aharonov-Bohm\cite{ABeffect}
effect: the optical interference of the light emitted from the two HJLs in the
SQUID depends periodically on the flux through the superconducting loop. In the
present study, we will make a step forward by including optical coupling in two
ways: (i) the HJLs share a single resonant mode, and (ii) the separate resonant
modes of the HJLs are coupled and (partly) hybridized. We will call these setups
light-superconducting interference devices (LSID).

The paper is organized as follows. In an introductory Sec.~\ref{sec:regularhjl},
we explain the essential results and equations for a single HJL. We introduce
two SQUID-based setups in Sec.~\ref{sec:setups}. These two setups are treated in
Secs~\ref{sec:singlemodedevice} and \ref{sec:dualmodeHJL} respectively. The
second setup provides a special regime where we find time-dependent periodic
lasing solutions. This is investigated in a separate Sec.~\ref{sec:limitcycle}.
We conclude in Sec.~\ref{sec:conclusion}.

\section{Introduction: The half-Josephson laser}\label{sec:regularhjl}
A single HJL is described in detail in Refs \onlinecite{godschalk,godschalk2}.
Here, we will give main equations and results for a simple but general model of
a multi-emitter HJL, formulated in Ref.~\onlinecite{godschalk2}. This is also
useful because of many similarities of the LSID with a single HJL with respect
to the classification of lasing regimes.

The half-Josephson laser can be regarded as a superconductor - {\it p-n} diode -
superconductor heterostructure mounted in an optical resonator. The {\it p-n}
diode is capable of emitting light by electron-hole
recombination\cite{recher:10}. In the model of Ref.~\onlinecite{godschalk2}, the
optical resonator mode is driven by a large number of quantum emitters, that
form a dipole moment oscillating at about half the Josephson frequency,
$\omega_j/2= eV/\hbar$, with $V$ the bias voltage. It is essential that
optically active eigenstates of the quantum emitters couple to the two
superconducting leads. This coupling then results in a phase lock between the
optical phase of the electric field in the resonator mode and the
superconducting phase difference between the leads. With increasing field
intensity in the mode, the dipole moment saturates, so that steady state lasing
occurs at finite field intensity. In a toy model, the HJL is driven by an {\it
a.c.} Josephson current with frequency $\omega_j$. The lasing in the HLJ occurs
as a result of a parametric resonance instability at $\omega_j/2$. Due to this,
there are two stable lasing states with optical phases shifted by $\pi$.  

Fluctuations in the laser intensity and phase of the HJL, originate from quantum
noise in the optical mode\cite{scullylamb} and spontaneous switchings between
eigenstates of the quantum emitters. Such fluctuations can lead to spontaneous
switchings between two stable lasing states and result in loss of optical
coherence. However, we have shown that the typical switching times can be
exponentially long\cite{godschalk2}. Therefore, in this work, we consider
neither noise nor switching in the devices under consideration. 

In Ref.~\onlinecite{godschalk2}, we have derived a general model for the HJL
applying to an arbitrary set of quantum emitters. The states of these quantum
emitters were assumed to couple only weakly to both the optical field and the
superconducting leads. This allowed us to express the dipole moment in terms of
an expansion in the optical field of the resonant mode and the pair potentials
of the superconducting leads. With the optical
field represented by the expectation value of the photon annihilation operator,
$b\equiv \langle\hat b \rangle$, its semiclassical equation of motion is given
by
	\begin{align}\label{eq:HJLeqs}
	   \dot b = -\left( i\omega + \frac{\Gamma}{2} \right)b 
			- i\Omega''|b|^2 b -i A b^*e^{i\phi_\Delta}.
	\end{align}
Here, $\omega$ is the detuning from the  frequency of the resonant
mode $\omega_0$, $\omega \equiv \omega_0 - eV/\hbar$, $\Gamma$ is the decay rate
of the mode and $\phi_\Delta$ the
superconducting phase difference. The coefficients $\Omega''$, and $A$
correspond to the third order terms in the expansion of the dipole moment. The
lowest order term is proportional to $b$ and shifts the resonant
frequency of the mode. The second order terms of the expansion are zero.
A coherent state of radiation is formed in the resonant mode, with the average
photon number being given by $n=|b|^2$. The equations are similar to generic
equations describing parametric resonance in the presence of a weak
non-linearity\cite{parametricresonance}. The superconductivity plays the role of
an a.c. drive at double frequency $2 eV/\hbar$. 

The stationary solutions to Eq.~\eqref{eq:HJLeqs} are given by $n=0$ and 
	\begin{align}\label{eq:HJLsol}
		\begin{split}
		n_\pm &= \frac{1}{|\Omega''|}\left[\pm\sqrt{A^2-\Gamma^2/4} + \omega 	
			\right], \\
		\frac{\Gamma}{2}\tan&\left(\varphi_b-\frac{\phi_\Delta}{2}\right) = -A\mp 
		\sqrt{A^2-\Gamma^2/4}, 
		\end{split}
	\end{align}
where we have assumed $\Omega''<0$. Here, $\varphi_b$ is the optical phase of
the field in the mode. The fixed value of the optical phase implies a phase lock
to
the superconducting phase difference. The solution for the phase is covariant
under $\varphi_b\to \varphi_b + \pi$, which implies the occurrence of two
solutions for each of the $n_\pm$, with opposite field amplitudes. 

To realize lasing in the HJL, at least one of the solutions $n_\pm$ must be real
and positive. This condition allows us to distinguish three regimes, depending
on the number of physical solutions. (i) Both $n_\pm$ are negative [case (i)a]
or
complex [case (i)b; here $n_+ = n_-^*$]. The only
physical solution to Eq.~\eqref{eq:singlecavity} is at $n=0$. (ii) Only
$n_+$ is real and positive. There are now two physical solutions, of wich the
one at $n=0$ is unstable against perturbations. This is the regime where we have
stable, steady state lasing with $n_+$ photons in the mode. To have a large
number of photons, it is required that
$|\Omega''|\ll\sqrt{A^2-\Gamma^2/4}+\omega$. (iii) Both $n_\pm$ are real and
positive, so that there are three physical solutions. Stability analysis shows
that only the solution with $n_-$ photons in the resonator mode is unstable.
Hence this regime is bistable, with both the nonlasing state and the lasing
state ($n_+$) stable against perturbations. 

In a phase diagram of $2A/\Gamma$ versus $2\omega/\Gamma$, regime (i)b borders
regimes (i)a and (iii). The boundary is simply defined by $A=\Gamma/2$.
Furthermore, regime (i)a borders (ii) and (ii) borders (iii). Here, the
boundaries are respectively given by $\pm |\omega| = \sqrt{A^2-\Gamma^2/4}$.

In a steady lasing state a constant current runs through the HJL. Since each
photon that escapes the resonator is replenished by an electron-hole pair
annihilation, the current is given by the number of photons that
escapes the cavity, $\Gamma n$, times the electric charge, $I=e\Gamma n$.

\section{setups}\label{sec:setups}
\begin{figure}
  \includegraphics{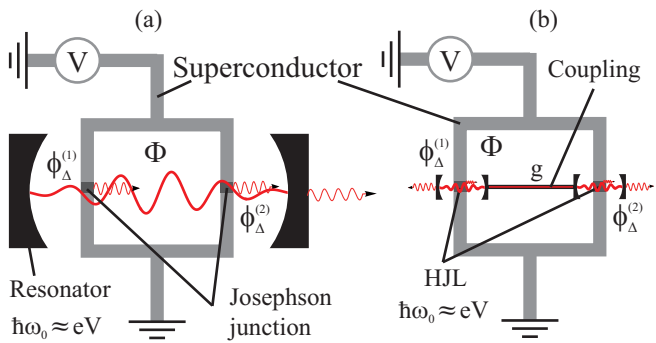}
  \caption{% 
	(Color Online) Setups. (a) The single mode LSID. Two HJLs, sharing the same
optical cavity, are embedded in a superconducting loop. The resonant frequency
is $\omega_0 \approx eV/\hbar$. 
(b)	The dual mode LSID: a superconducting loop containing a HJL in each
arm. The HJLs are embedded in separate cavities with resonant mode frequencies
$\omega_{1,2} \approx eV/\hbar$. The optical coupling between the resonant modes
is characterized with a parameter $g$ , such that the splitting of the
frequencies of the hybridized modes equals $2g$.
  }\label{fig:setups}
\end{figure}
Let us introduce two LSID setups and the corresponding equations of
motion for optical fields.

The first setup contains two HJLs sharing a single cavity, which are embedded
in the arms of a superconducting loop. This is similar to the design of a {\it
d.c.} SQUID. A magnetic flux $\Phi$ threads the loop of the SQUID-structure,
thus relating the superconducting phase
differences across the Josephson junctions, defined by $\phi_\Delta^{(j)}$ for
$j=1,2$, so that $\phi_\Delta^{(1)}-\phi_\Delta^{(2)} = 2\pi\Phi/\Phi_0$. Here
$\Phi_0=\pi\hbar /e$ is the quantum of magnetic flux. 
From now, we refer to this setup as `single mode LSID'.

The description of the single mode LSID is based on the phenomenological
model that was introduced with Eq.~\eqref{eq:HJLeqs} for a single HJL. We
account for two HJLs  by splitting the term proportional to $b^*$. Each part
comes with its own coefficient, $A_j$, $j=1,2$, and the corresponding
superconducting phase difference, $\phi_\Delta^{(j)}$. Redefining the optical
phase of $b$, $\varphi_b \to\varphi_b + \phi_\Delta^{(1)}/2$, we arive at
	\begin{align}\label{eq:singlecavity}
	   \dot b &= -\left( i\omega + \frac{\Gamma}{2} \right)b 
			- i\Omega''|b|^2 b\\
			&-i A_1 b^* -i A_2 b^* 
			e^{-2i\pi\Phi/\Phi_0}. \nonumber
	\end{align}
Here, $\omega$, $\Gamma$ and $\Omega''$ are the same as in
eq.~\eqref{eq:HJLeqs}, while $A_1$ and $A_2$ are equivalent to $A$ in the model
of a single HJL. Also here, without loss of generality, we assume from now on
$\Omega''<0$. In case of similar HJLs in the arms of the SQUID, $A_1 \simeq
A_2$. We note that the equation of motion for a single HJL is obtained by
setting $\Phi=0$. 

The second setup is similar to the first one, with the exception that there are
now two resonant modes, each associated with a HJL. The modes are coupled
optically. For instance, this can be realized if each HJL is mounted in a
separate optical cavity, the cavities being connected with a fiber. Also here, a
flux $\Phi$ threads the loop. This device will be referred to as
`two-mode LSID'.

We model this setup using two copies of the equations of motion for a single
mode HJL, Eq.~\eqref{eq:HJLeqs}, and by augmenting those with a coupling
term\cite{haus}. Assuming $b_j$ to be the optical field in the modes labelled
with $j=1,2$, we arrive at 
	\begin{align}\label{eq:doublecavity}
	   \dot b_1 &= -\left( i\omega_1 + \frac{\Gamma_1}{2} \right)b_1 
			- i\Omega_1''|b_1|^2 b_1   \nonumber \\
			&-i A_1 b_1^*  -i g b_2 
			e^{-i\pi\Phi/\Phi_0}, \\
	   \dot b_2 &= -\left( i\omega_2 + \frac{\Gamma_2}{2} \right)b_2 
			- i\Omega_2''|b_2|^2 b_2  \nonumber \\
			&-i A_2 b_2^*	-i g b_1 e^{i\pi\Phi/\Phi_0}.
			\nonumber
	\end{align}
Here, $\omega_j$ is the detuning of each mode and $\Gamma_j$ the decay rate. The
coefficients $\Omega_j''$ and $A_j$ are coefficients of the expansion of the
dipole moments. Like for the model of the single mode, we redefined here the
optical phases: $\phi_b^{(j)}\to\phi_b^{(j)} + \phi_\Delta^{(j)}/2$. The
coupling between the modes is proportional to coupling strength $g$, which we
take real, without loss of generality.

It is worth noting that, compared to the first setup, the second setup has more
output quantities: one can separately measure intensity and optical phase of the
light emitted from each mode. 

Schematics of both setups are shown in Fig.~\ref{fig:setups}.

\section{Single mode LSID}\label{sec:singlemodedevice}
Let us now analyse the model of the single mode LSID.
The stationary solutions to Eq.~\eqref{eq:singlecavity} yield the stationary
number of photons in the resonator mode, $n=|b|^2$, and the optical phase. They
are given by
	\begin{align} \label{eq:solssinglemode}
	   |\Omega''|n_\pm &= \pm \sqrt{A_1^2+A_2^2+2A_1A_2 	
	   	\cos\left(2\pi\frac{\Phi}{\Phi_0}\right) -
		\frac{\Gamma^2}{4}} + \omega, \nonumber
		\\
		\tan2\varphi_b^\pm &= \frac{\frac{\Gamma}{2}\big[A_1 + 		
			A_2\cos\big(2\pi\frac{\Phi}{\Phi_0}\big)\big] \mp 
			W A_2 \sin\big(2\pi\frac{\Phi}{\Phi_0}\big)}{\pm W\big[A_1 + 
			A_2\cos\big(2\pi\frac{\Phi}{\Phi_0}\big)\big] + 
			\frac{\Gamma}{2}A_2 \sin\big(2\pi\frac{\Phi}{\Phi_0}\big)	},
	\end{align}
with $W = \omega -|\Omega''|n_\pm$. Besides, $n=0$ is also a stationary
solution. The expression for $\varphi_b$ implies that a stationary state with
photon number $n_\pm$ can occur with two phases, differing by $\pi$.
Furthermore, the physical solutions correspond to real and positive $n_\pm$. As
a minimal requirement for lasing, we need $|A_1+A_2|>\Gamma/2$. From now on, we
assume this to be the case. It is essential to note that the $n_\pm$ depend on
the magnetic flux. In
particular, the threshold values of $\omega$ at which $n_\pm=0$, depend
periodically on
$\Phi$: $\omega_\rs{thr}^\pm(\Phi)$. The sensitivity to flux is highest when
$|A_1-A_2|<\Gamma/2$. In this case, there is a value of $\Phi$, where the
expression in the square root becomes zero, so that $n_-=n_+ =
\omega/|\Omega''|$.

\begin{figure}
  \includegraphics{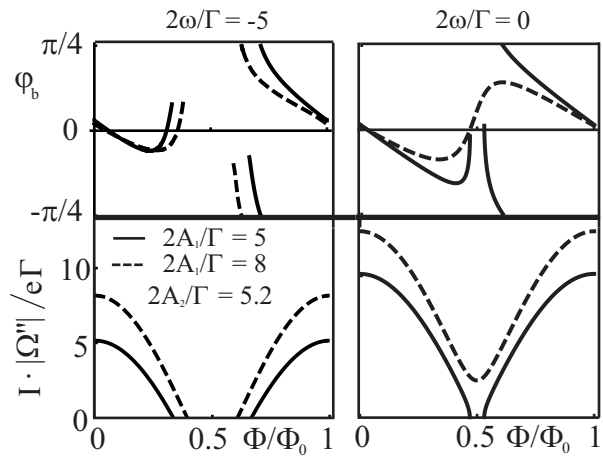}
  \caption{% 
	Flux dependence of the current and optical phase, and regime switching for
the single mode LSID. In the lower panels, the
current through the device is plotted as a function of the flux, $\Phi$. The
current is proportional to the $n_+$ given by Eq.~\eqref{eq:solssinglemode}.
In the upper panels, the corresponding optical
phases are plotted. For the solid (dashed) lines $|A_1-A_2|<\Gamma/2$
($|A_1-A_2|>\Gamma/2$). For the leftmost panels, the detuning is chosen such
the HJL undergoes a transition between the regimes (ii) and (i)a described
in the main text. This occurs near the half of a flux quantum. This happens for
the both cases corresponding to the dashed and the solid lines. For the
rightmost panels, no
transition occurs for the dashed line, while a transition from
regime (ii) to (i)b occurs for the solid line, in the vicinity of
a half of flux quantum. 
  }\label{fig:currentsinglecavity}
\end{figure}

The single mode LSID can operate in three different regimes\cite{godschalk2}.
The definition of these regimes is exactly the same as that of those of the
single HJL, while the boundaries separating the regimes are different and the
dimensionality of the phase diagram is higher, involving the extra parameters
$A_2$ and $\Phi$. 

It is possible to switch between the regimes by changing parameters. For
instance, for a single HJL (equivalent to setting $\Phi=0$ for the single mode
LSID) one can switch from regime (i) to (ii) and from (ii) to (iii) by
sweeping the voltage and thus the detuning $\omega$. With the single mode LSID,
new possibilities arise to switch between the regimes. A very interesting one is
a switch between regimes (i) and (ii),
a nonlasing and a lasing regime, by changing the flux only. This can happen in
two ways. First, we can choose $\omega$ such that it is crossed by
$\omega_\rs{thr}^+(\Phi_\rs{thr}^{+,(i)})$ at the threshold value of the
flux, $\Phi_\rs{thr}^{+,(i)}$. This is a transition between regimes (i)a and
(ii). It corresponds to a second order phase transition, where the derivative of
$n$ to $\Phi$ is finite when the threshold is crossed. The second way occurs
when $|A_1-A_2|<\Gamma/2$
and $\omega=0$. Here a second order phase transition between regimes (i)b and
(ii) occurs at the two threshold values of $\Phi$. At this phase transition we
find $n_- = n_+=0$, while the derivative of $n$ to $\Phi$ is infinite. These
cases are shown in fig.~\ref{fig:currentsinglecavity}, where the current through
the device and the optical phase are plotted as a function of flux. Hence, with
these phase transitions it is possible to switch a single mode LSID on and off
using a magnetic field only.

\begin{figure}
  \includegraphics{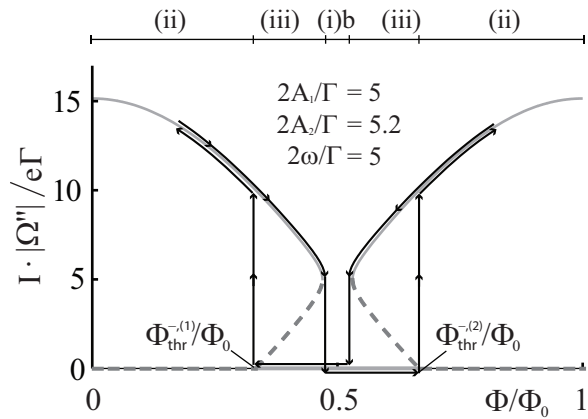}
  \caption{% 
	The hysteresis in the single mode LSID. The bistable regime (iii) supports
hysteretic
behaviour in the HJL. The gray, solid (dashed) curve represents the current as
calculated from the stable (unstable) solution of
Eq.~\eqref{eq:solssinglemode}. The solid (dashed) line at $I=0$ indicates that
that the non-lasing solution is stable (unstable). The solid black lines
represent a flux sweep, with the direction indicated by the arrows. These lines
are slightly shifted for clarity. The regimes (i)b, (ii) and (iii), indicated
above the plot, and the threshold flux values, $\Phi_\rs{thr}^{-,(i)}$, are
explained in the main text.
  }\label{fig:hysteresis}
\end{figure}

There is also a parameter regime where the single mode LSID displays
hysteretic behaviour upon a flux sweep. This regime occurs when
$|A_1-A_2|<\Gamma/2$ and $\omega$ is chosen such that
$\omega_\rs{thr}^-(\Phi_\rs{thr}^{-,(i)})=\omega$ (the `threshold' of the
unstable solution), at the threshold value of the flux, $\Phi_\rs{thr}^{-,(i)}$.
If we start at $\Phi=0$, the HJL is in regime (ii). Upon increasing $\Phi$
adiabatically, a transition to the bistable regime (iii)
takes place when the threshold $\Phi_\rs{thr}^{-,(1)}$ is crossed. The single
mode LSID remains in the steady lasing state. At a critical value of $\Phi$ we
encounter a transition to the non-lasing regime (i)b. This is a first order
phase transition, where the single mode LSID turns off. When we  decrease
$\Phi$, the transition proceeds in opposite direction, from regime (i)b to
(iii). Since the non-lasing state is stable in this regime, the HJL remains off.
Crossing $\Phi_\rs{thr}^{-,(1)}$ another time, we encounter a first order
transition to the original lasing regime (ii). The hysteresis in the HJL is
shown in fig.~\ref{fig:hysteresis}, where the sweep occurs over a wider range of
$\Phi$, which also includes a second threshold, $\Phi_\rs{thr}^{-,(2)}$.

To conclude this section, we have described a single mode LSID, where two HJLs
sharing the same resonant mode are incorporated in a superconducting loop. We
find the single mode LSID to be flux-tunable.
Importantly, in some parameter regimes the lasing can even be switched on and
off solely using the small magnetic fields. Additionally, a parameter regime
exists where there is a hysteresis with respect to a flux sweep.

\section{Two-mode LSID}\label{sec:dualmodeHJL}
In this section we analyse the model, Eq.~\eqref{eq:doublecavity}, of the
two-mode LSID, where a superconducting loop contains a HJL in each arm of the
loop, while the resonant modes are coupled optically. The relative complexity of
this model prohibits us from doing a full analytical study. Instead, we
investigate the weak and the strong coupling limits using perturbative methods. 
Then we study analytically the equations for a specific, symmetric
choice of parameters, assuming no particular coupling strength. For a
particular parameter range of the latter case, we also perform a numerical
study, in Sec.~\ref{sec:limitcycle}, where we find time-dependent solutions to
Eq.~\eqref{eq:doublecavity}.

\subsection{Weak coupling limit}
Let us first study the weak coupling limit, $g\ll A_i$, of the two-mode
LSID. In this limit, the two HJLs in the device only slightly perturb each
other. The perturbation depends on the flux, $\Phi$. As a result, the current
through the device displays small oscillations upon changing flux.

We calculate the flux dependent change in the optical fields of the modes in the
weak coupling limit. The stationary lasing states of the uncoupled HJLs, given
in Eq.~\eqref{eq:HJLsol}, are taken to be $n_j^0=|b_j^0|^2$ and
$\varphi_{b_j}^0$, with the index $j=1,2$ labelling the HJLs. For clarity, we
make an extra assumption $\Gamma_j\ll A_j$ and expand Eq.~\eqref{eq:HJLsol}
about
$\Gamma_j=0$. In this limit, $(x_j^0,y_j^0) =
(\sqrt{(A_j+\omega_j)/|\Omega_j''|},0)$. The optical coupling in
Eq.~\eqref{eq:doublecavity} and the $\Gamma_j$ perturb the optical fields as 
$b_j = b_j^0+\delta
b_j \equiv x_j^0+iy_j^0 + \delta x_j + i \delta y_j$, with $x_j^0,y_j^0 , \delta
x_j , \delta y_j$ being real. We calculate the linear variations owing to
$\Gamma_j$ and $g$, which yields
	\begin{align}
		\begin{split}
		\delta^\Gamma x_j &= 0,\\
		\delta^\Gamma y_j &= -x_j^0\
			\frac{\Gamma_j}{2A_j}, \\
		\delta^g x_j &= x_k^0\
			\frac{g}{2 (A_j+\omega_j)} \cos[\pi\Phi/\Phi_0], \\
		\delta^g y_j &= x_k^0\ (-1)^j
			\frac{g}{2 A_j} \sin[\pi\Phi/\Phi_0],
		\end{split}
	\end{align}
where $j,k=1,2$ and $j\ne k$. 

As a result of the perturbative interaction, the current through the HJL and the
optical phase change. The total
current through the device becomes $I \simeq I_0+ \delta I$, with $I_0 = e\sum_j
\Gamma_j n_j^0$ and $\delta I = 2e\sum\Gamma_i(x_i^0 \delta x_i + y_i^0 \delta
y_i)$. Here $I_0\simeq \sqrt{n_0} \delta I$. The variations owing to $\Gamma_j$
yield a small constant reduction of the total current in second
order, while the ones owing to the optical coupling yield, in first order, a
small flux dependent change of the current. The perturbation to the
optical phase is given by $\delta \varphi_{b_j} = \cos^2[\pi\Phi/\Phi_0](\delta
y_j^0- \tan\varphi_{b_j}^0\delta x_j^0)/x_j^0$. Up to first order, we find
	\begin{align}
	   \delta I &= e g\left[\frac{\Gamma_1}{|\Omega_1''|} 
		\sqrt{\frac{n_2^0}{n_1^0}} + \frac{\Gamma_2}{|\Omega_2''|} 
		\sqrt{\frac{n_1^0}{n_2^0}} \right] \cos[\pi\Phi/\Phi_0], \\
		\delta \varphi_{b_j} &= \frac{\cos^2[\pi\Phi/\Phi_0]}{2A_j} 
		\left[-\Gamma_j + (-1)^j g \sqrt{\frac{n_k^0}{n_j^0}} \sin[\pi\Phi/\Phi_0]
		\right], \nonumber
	\end{align}
where $k\ne j$ and $n_j^0 = (x_j^0)^2 = (A_j+\omega_j)/|\Omega_j''|$. The phase
variation can be written as a sum of simple harmonic functions, with
arguments $m\pi\Phi/\Phi_0$, for $m=1,2,3$. 

\subsection{Strong coupling limit}
We proceed with the strong coupling limit of the two-mode LSID assuming $g\gg
A_i,\Gamma_i$. In this limit, the modes of the HJLs are essentially hybridized.
The frequencies of the hybridized modes are shifted by $\pm g$. We show that
each of these hybridized modes is excited separately in separate ranges of
detuning. In these ranges, the two-mode LSID works similar to a single mode
LSID. 

A perturbative treatment of Eq.~\eqref{eq:doublecavity}, requires tuning to one
of the two hybridized modes, $\omega_i\simeq\pm g$. With this, the lowest order
stationary solution reads $\omega_2 b_2^{(0)} = -g b_1e^{i\pi\Phi/\Phi_0}$.
Then, up to first order we have
$\pm\omega_2 b_2 = [\pm\omega_2 +i\Gamma_2 - |\Omega_2''| n_2^{(0)}] b_2^{(0)} -
A_2[b_2^{(0)}]^*$. Inserting these results in the expression for $b_1$, yields
the equation for a stationary single mode LSID
	\begin{align}
	   &\left[ \pm i\left( \omega_1- \frac{g^2} {\omega_2^2 }\omega_2 \right) 
			+\frac{\Gamma_1}{2} + \frac{\Gamma_2}{2}\frac{g^2} {\omega_2^2 } 
		\right] b_1 =  \\
		&-i \left( \Omega_1'' + \frac{g^4} {\omega_2^4 }\Omega_2'' \right) n_1
			 b_1
		- i\left( A_1 + \frac{g^2} {\omega_2^2 }A_2 e^{-2i\pi\Phi/\Phi_0} \right) 
		b_1^*. \nonumber
	\end{align}
The approximation leading to this equation is valid for a limited range of
detunings, $|\omega_1-(g^2/\omega_2)|\lesssim |A_j|$.

\subsection{Symmetric equations}
The limits studied so far give a rather narrow perspective of the two-mode LSID:
in the weak coupling limit it is described as two largely
independent HJLs, while in the strong coupling limit it essentially becomes a
single mode LSID, at least for a narrow interval of detuning.  To learn more
about the device, let us assume the HJLs to be identical. With this, it is
possible to
analytically calculate stationary solutions to Eq.~\eqref{eq:doublecavity}.
Small deviations from this assumption of symmetry can in principal be treated
perturbatively. Doing so, we have not found any qualitative differences from the
symmetric case. Hence, we describe all essentials of the two-mode LSID for
the case when the arms of the superconducting loop contain equal HJLs. 

Before presenting the stationary solutions, we first reduce the parameter space
of Eq.~\eqref{eq:doublecavity} by rescaling various quantities to dimensionless
form
	\begin{align}\label{eq:scaling}
	\begin{split}
	   \tilde b_j &\equiv \sqrt{\frac{|\Omega_j''|}{A_j}}\ b_j, 
	   \quad \gamma_j \equiv \Gamma_j/(2A_j),  \\
		G_{jk} &\equiv \frac{g}{A_j}\sqrt{ \frac{|\Omega_j''|}{|\Omega_k''|} 
			\frac{A_k}{A_j}}, \quad 
	   \tilde\omega_j \equiv \omega_j/A_j, 
		\end{split}
	\end{align}
and measuring time in units of $(A_1 A_2)^{-1/2}$.
With this, the equations of motion become
\begin{align}\label{eq:doublecavityreduced}
	   \sqrt{\frac{A_2}{A_1}}\dot{\tilde b}_1 &= -\left( i\tilde\omega_1 +
\gamma_1 \right)\tilde b_1 
			+ i|\tilde b_1|^2 \tilde b_1   \nonumber \\
			&-i \tilde b_1^*  -i G_{12} \tilde b_2 
			e^{-i\pi\Phi/\Phi_0}, \\
	   \sqrt{\frac{A_1}{A_2}}\dot{\tilde b}_2 &= -\left( i\tilde \omega_2 +
\gamma_2 \right)\tilde b_2 
			+ i|\tilde b_2|^2 \tilde b_2  \nonumber \\
			&-i \tilde b_2^*	-i G_{21} \tilde b_1 e^{i\pi\Phi/\Phi_0}.
			\nonumber
	\end{align}
The assumption of symmetry implies $\tilde\omega_1=\tilde\omega_2
=\tilde\omega$, $\gamma_1=\gamma_2\equiv \gamma$, and $G\equiv G_{12}=G_{21}$
(we note that this does not imply $A_1=A_2$). For this choice, the {\it
stationary} solutions of the equations of motion are
invariant under exchange of the resonators $(1\leftrightarrow 2)$ and
reversing the magnetic field $\Phi\to-\Phi$. Because of this $\tilde
n\equiv\tilde n_1 = \tilde n_2$, with $\tilde n_j = |\tilde b_j|^2$, while
$\varphi_{b_1}\ne \varphi_{b_2}$.

We have found five stationary solutions, either stable or unstable, to
Eqs~\eqref{eq:doublecavityreduced} in the symmetric case, for photon number and
optical phase. This includes $n=0$. For brevity, we give the expression of the
optical phases only in the limit $\gamma \to 0$ 
	\begin{align}
		&\tilde n_\alpha^\pm = \pm\sqrt{1+G^2-\gamma^2 + 2G 
		\sqrt{\cos^2\left(\pi\frac{\Phi}{\Phi_0}\right) - \gamma^2}}
			+ \tilde\omega, \nonumber \\
			&2\varphi_{b,\alpha}^{(1)} +\frac{\pi\pm\pi}{2} = \arctan\left[\frac{-G
		\sin \left(\pi\frac{\Phi}{\Phi_0}\right)}
		{1+G\cos\left(\pi\frac{\Phi}{\Phi_0}\right)}\right] \equiv \varphi_G^+, 
				\nonumber \\
			&\text{with }\, \varphi_{b,\alpha}^{(1)}=-\varphi_{b,\alpha}^{(2)},
		\label{eq:soldoublecavityalpha}\\
		&\tilde n_\beta^\pm = \pm\sqrt{1+G^2-\gamma^2 - 2G 
		\sqrt{\cos^2\left(\pi\frac{\Phi}{\Phi_0}\right) - \gamma^2}} +
		\tilde\omega, \nonumber \\
		&2\varphi_{b,\beta}^{(1)} \mp \frac{\pi}{2}
		 = \arctan\left[ \frac{G \sin \left(\pi\frac{\Phi}{\Phi_0}\right)}
		{1-G\cos\left(\pi\frac{\Phi}{\Phi_0}\right)} \right] \equiv \varphi_G^-, 
		\nonumber	\\
		&\text{with }\,\varphi_{b,\beta}^{(1)} 
		=-\varphi_{b,\beta}^{(2)}-\frac{\pi}{2}.
			\label{eq:soldoublecavitybeta}
	\end{align}
Let us make several remarks. First, the solutions are invariant under a change
of both
optical phases with $\pi$. This is equivalent to the invariance of
Eq.~\eqref{eq:doublecavityreduced} under a sign change of both $\tilde b_1$ and
$\tilde b_2$. Second, the solutions are periodic in flux, with the flux period
of $2\Phi_0$. This period is however only visible in the dependence of the
optical phases on the flux, that can be probed by measuring the light
interference. In contrast, the current through the device is only sensitive to
the photon number, which has a flux period of $\Phi_0$. Finally, for $\gamma\ne
0$ there is a region of flux values, defined by $\gamma>|\cos[\pi\Phi/\Phi_0]|$,
where the $\tilde n_{\alpha,\beta}^\pm$ are complex valued, so that the only
physical solution is at $\tilde n=0$. This regime is similar to the regime (i)b
that was discussed in context of the single mode LSID in
Sec.~\ref{sec:singlemodedevice}.

\begin{figure*}
  \includegraphics{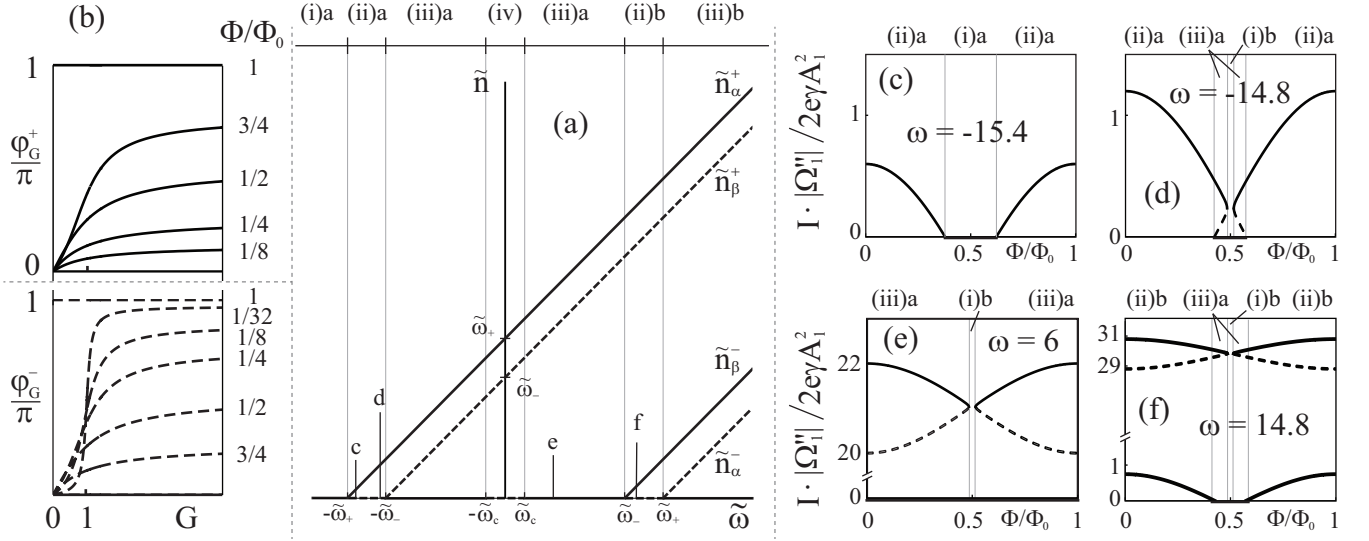}
  \caption{%
  The lasing in the two-mode LSID. (a) The four solutions of $n$ as given by
eqs~\eqref{eq:soldoublecavityalpha} and~\eqref{eq:soldoublecavitybeta}, as a
function of detuning. This interval of $\tilde \omega$ corresponds to various
regimes described in the main text. The regimes are indicated by labels above
the plot. The values of detuning labelled with c-f correspond to the panels
(c)-(f) at $\Phi=\Phi_0/4$. In the strong coupling limit, the solid lines
labelled with $\tilde n_\alpha^+$ and $\tilde n_\beta^-$ are the stable
stationary solutions, while the dashed lines labelled with $\tilde n_\beta^+$
and $\tilde n_\alpha^-$ are the unstable ones. 
We have defined $\tilde\omega_\pm^2\equiv 1+G^2-\gamma^2 \pm 2G
\sqrt{\cos^2[\pi\Phi/\Phi_0] - \gamma^2}$. In the limit of $G\gg1$, the critical
value of the detuning for regime (iv) is given by $\tilde\omega_c =
\sqrt{\sin^2[\pi\Phi/\Phi_0] - \gamma^2}$ (Sec.~\ref{sec:limitcycle}).
(b) The optical phases as a function of $G$ for several values of the flux, as
explained in the main text. The solid (dashed) curves correspond to the solid
(dashed) curves in (a), for $0<\Phi \leq\Phi_0$.
(c)-(f) The stationary current (thick solid curves) as a function of flux, for
several values of the detuning. The dashed curves correspond to the current at
the unstable stationary solutions. Panels (d)-(f) contain bistable regimes where
two values of the stationary current are possible. With changing flux, switches
between various regimes, indicated above the panels, occur. These panels
correspond to the strong coupling limit, with $G=15$. Furthermore, $\gamma=0.05$
and $|\Omega_1''|A_2^2 = |\Omega_2''|A_1^2$.
  }\label{fig:doublecavity}
\end{figure*}

Figure~\ref{fig:doublecavity} presents the plots of the solutions of Eqs
\eqref{eq:soldoublecavityalpha} and \eqref{eq:soldoublecavitybeta}.
The solutions $n_{\alpha,\beta}^\pm$ are shown in
Fig.~\ref{fig:doublecavity}(a) (plot in the center). In this panel, we can
distinguish
the various regimes that occur in this device, those are similar to the ones
introduced for the single mode LSID in Sec.~\ref{sec:singlemodedevice}. In
regime (i)a no lasing occurs in the two-mode LSID. Regime (i)b is not shown in
the plot, while it was mentioned in the previous paragraph. The lasing occurs in
the regime (ii). There is a single stable lasing solution in (ii)a and there are
two stable lasing solutions in (ii)b. The latter also involves an unstable
lasing solution. Regime (iii)a is bistable while (iii)b is tristable. These
regimes also contain one and two unstable solutions respectively. In both cases,
the non-lasing solution is stable. Finally, in the vicinity of $\tilde \omega=0$
there is a new regime (iv). This regime contains time-dependent solutions (limit
cycles) and will be the topic of investigation in Sec.~\ref{sec:limitcycle}.
 
We see that steady state lasing occurs in regime (ii), in two small windows of
the detuning, those being in the vicinity of $\tilde\omega=\pm G$. Hence, indeed
as expected, we find two lasing modes at a frequency shifted by the coupling
constant $\simeq G$ and a frequency splitting of $\simeq 2G$.

The stability of the solutions found depends on the coupling strength. For
$G\gg1$, 
the solid lines ($\tilde n_\alpha^+$ and $\tilde n_\beta^-$) in
fig.~\ref{fig:doublecavity}(a) represent the stable solutions. The dashed lines
represent to unstable ones. In the limit $G\ll 1$ we find that $\tilde
n_\beta^+$ is stable instead of
unstable, while $\tilde n_\beta^-$ is unstable in stead of stable. This
is expected in the regime where the two HJLs in the arms of the superconducting
loop are only coupled weakly. Here, both HJLs should lase in a regime of
detuning about $\tilde \omega =0$. The stable solutions merge at
$G\to0$, as do the unstable ones.

The dependence of the optical phases on the flux is shown in
Fig.~\ref{fig:doublecavity}(b). Instead of showing the value of each solution of
the phase separately, we have plotted $\varphi_G^\pm$. It is sufficient to plot
in a  flux interval 
from zero to $\Phi_0$. Indeed, $\varphi_G^\pm$ for $0<\Phi\leq\Phi_0$, is the
same
as $\varphi_G^\mp$ for $\Phi_0<\Phi\leq2\Phi_0$. At $G=0$ the phase
$\varphi_G^\pm$ is either zero or $\pi$ while for $G\to\infty$,
$\tan\varphi_G^\pm =\pm\tan[\pi\Phi/\Phi_0]$.

Finally, Figs~\ref{fig:doublecavity}(c)-(f) show the (possible) stationary
currents as a function of the flux in the strong coupling limit. 
Similar to the single mode LSID, the flux can be used to change the operating
regime of the device.
Panels (c) and (d) correspond to
the same regimes as the left and right panel (for the latter only the solid
line) of Fig.~\ref{fig:currentsinglecavity} respectively. Indeed, a flux sweep
in the parameter regime of panel (d), would show hysteresis, similar to what is
shown in fig.~\ref{fig:hysteresis} in Sec.~\ref{sec:singlemodedevice}. In panel
(e) a regime change between the bistable regime (iii)a and the nonlasing regime
(i)b occurs. If in the lasing state, a flux sweep across the point
$\Phi_0/2$ extinguishes the lasing  without recovering. Panel (f) shows
transitions between regimes (ii)b, (iii)a and (i)b.

To conclude this section, we have studied the two-mode LSID. In the weak
coupling limit, the effect of the flux is small
periodic modulations at the background of the current for two uncoupled HJLs. In
the strong coupling limit, there are intervals of the detuning where the device
operates like the single mode LSID while the overall picture is more complex. In
the next section, we concentrate on  a nontrivial feature that is unique for the
two-mode LSID.

\section{Periodic lasing cycles}\label{sec:limitcycle}
In the previous sections, we have studied the stationary states of the
LSID. As noted, the two-mode LSID displays a
regime with time-dependent steady  solutions, or `limit cycles'. These
are the topic of this section. First, we give a theoretical background to this
phenomenon. We do stability analysis to find the parameter ranges where this
interesting regime takes place, and identify the corresponding dynamics of the
LSID. Then, we use a perturbative analysis in the limit of strong coupling $G\gg
1$ to estimate the key properties of the limit cycles. We find that in this
limit, the emission predominantly occurs at two frequencies separated from
$eV/\hbar$ by $\pm g$. We refer to this as dual mode lasing. 
 
After this, we present the results based on the numerical
integration of the differential equation and compare these with the theoretical
estimates.  
 
\subsection{Stability}
We study the stability of the non-lasing solution, $\tilde b_j=0$, in the
vicinity of $\tilde\omega=0$. As in the previous section, we assume equal
parameters $\tilde\omega\equiv \tilde\omega_{1,2}$, $\gamma\equiv \gamma_{1,2}$,
$G_{12}=G_{21}$. In addition, we assume $A_1=A_2$.
The eigenvalues of the linearized equations of motion in the vicinity $\tilde
b_j=0$
read
	\begin{align}
	  \lambda \equiv \gamma \pm \sqrt{ 1- G^2 - \tilde\omega^2 \pm 2G	
	  	\sqrt{\tilde\omega^2 -\sin^2[\pi\Phi/\Phi_0] } },
	\end{align}
for all four possible combinations of the `$\pm$'s. From this, we can
resolve the various regimes defined in Sec.~\ref{sec:dualmodeHJL}.
For instance, the solution at $n=0$ is stable when the real parts of all
$\lambda$ are positive. In the lasing regime (ii) all $\lambda$ are real, yet
three are positive and one is negative, thus indicating a saddle point
instability. In regime (iv), the nonlasing solution is also unstable, but here
the corresponding eigenvalues are complex in stead of real, while the real part
of two eigenvalues is negative. This regime can only occur if $\tilde\omega^2 <
\sin^2[\pi\Phi/\Phi_0]$. In the remainder of this section we will always assume
$G\gg1$, so that regimes (ii) and (iv) are clearly separated from each other.
Then $\lambda$ is approximated as
	\begin{align}
		\lambda\simeq \gamma\pm  \sqrt{\sin^2[\pi\Phi/\Phi_0] 
		-\tilde\omega^2 } \pm 
		i \left[ G + \frac{\tilde\omega^2 - 1}{2G}  \right].
	\end{align}
again for all four possible choices of the `$\pm$'s. Therefore, in this
limit, the threshold for regime (iv) is defined by $\gamma^2 +\tilde\omega^2 =
\sin^2[\pi\Phi/\Phi_0]$. Crossing this threshold corresponds to a transition
from regime (iii)a to regime (iv).

To understand the implications of the transition to regime (iv), let us first
consider briefly the dynamics of the two-mode LSID in regime (iii)a. We discuss
it in terms used in Sec. IV of Ref.~\onlinecite{godschalk2}. We assume that the
LSID is not in a stationary state. Then the evolution of the state of the device
is governed by eq.~\eqref{eq:doublecavityreduced}. The optical fields, $\tilde
b_j$, can be decomposed into real and imaginary parts, $\tilde b_j = x_j + i
y_j$. Using these, we can construct a four-dimensional coordinate space where 
each point, $(x_1,y_1,x_2,y_2)$, represents a state of the two-mode LSID.
We can map the state evolution to the motion in the coordinate space of a
``particle'' which is driven by a ``force field''. Given some inital condition,
the particle will evolve along a trajectory defined by the force field, to a
stable stationary point or `attractor'. The set of initial conditions from which
the particle flows to one specific attractor is the domain of attraction of that
attractor.

In contrast to the attractors, some stationary points are unstable saddle
points. Generally, when close to a saddle point, the particle will be repelled
by it. There are however trajectories, that lead the particle to the saddle
point without it being repelled. These trajectories form the stable direction of
the saddle point and form a separatrix of Eq.~\eqref{eq:doublecavityreduced}. In
the
cases of regimes (ii)a and (iii)a of the two-mode LSID, we have
respectively one and two saddle points, for which the separatrix is
three-dimensional. Therefore, in the regimes (i) - (iii) the separatrices of
$m-1$ saddle points divide the state space in $m$ domains of attraction, each
associated to a single attractor. Because trajectories of the particle with
different initial conditions do not cross, it is not possible to switch from one
region to another without accounting for noise\cite{godschalk2}.

In the course of a transition from regime (iii)a to (iv) the non-lasing solution
becomes unstable.  However, as we have seen, the unstable direction is
two-dimensional in
stead of one dimensional. It cannot separate the region of the former attractor
at $n=0$ in two new regions, each with their own attractor. Importantly, the
attractors (saddle points) represented by the solution $n_\alpha^+$
($n_\beta^+$) and the separatrices do not change significantly, and no new
stationary attractors appear. Paradoxically, a particle in the domain of
attraction of the former attractor at $n=0$, is not evolving to an attractor
anymore, but it also cannot escape to another domain of attraction or to
infinity. To resolve this issue, this domain must contain a non-stationary
attractor, or limit cycle.

If the frequency of the limit cycle is $\omega_c$, one generally expects the
emission to occur at a comb of frequencies separated by $\omega_c$, $\omega_n =
eV/\hbar + n \omega_c$. Below we consider the limit of strong coupling where the
emission predominantly occurs at two frequencies corresponding to $n=\pm 1$.

\subsection{Perturbative analysis}
In the limit of $G\gg1$, it is possible to perform a perturbative analysis of
the regime (iv).
We use the full time dependent Eq.~\eqref{eq:doublecavityreduced}. Here, we
perform this analysis only up to first order in $ G^{-1}$. The results of this
subsection explain key features of the numerical results presented in the next
subsection.

To analyse Eq.~\eqref{eq:doublecavityreduced} perturbatively, we expand the 
fields in a series of $G^{-1}$: $b_j = b_j^{(0)}+G^{-1} b_j^{(1)}$ assuming
typical timescales of the order of $G^{-1}$. The lowest order equations read
	\begin{align*}
		\dot{\tilde b}_1^{(0)} + iGe^{-i\pi\Phi/\Phi_0} \tilde b_2^{(0)}=0, \quad 
		\dot{\tilde b}_2^{(0)} + iGe^{i\pi\Phi/\Phi_0} \tilde b_1^{(0)}=0.
	\end{align*}
The solutions can be found straightforwardly as
	\begin{align*}
		\tilde b_1^{(0)}(t) = -i\beta e^{-i\pi\Phi/\Phi_0}\sin[Gt], \quad
		\tilde b_2^{(0)}(t) = \beta\cos[Gt],
	\end{align*}
where we have implicitly chosen an origin in time, $t_0$. The complex constant
$\beta$ has yet to be determined. It will be fixed by the requirement that the
part of $b_j$ oscillating with frequency $G$, can be fully contained in the
leading order that includes $\tilde b_1^{(0)}(t)$ and $\tilde b_2^{(0)}(t)$. The
higher order
terms in the expansion only oscillate with frequencies that are multiples of
$G$. The time average of the total number of photons is $|\beta|^2 = \tilde
n_1^{(0)} + \tilde n_2^{(0)}$. This quantity is also proportional to the average
current through the device.

We continue with the first order corrections. To find these, we first take the
time derivative of Eq.~\eqref{eq:doublecavityreduced} and then collect all terms
that are proportional to $G$. To this end, we keep in mind that each time
derivative adds a factor of $G$. We find
	\begin{align}\label{eq:LCfirstorderG}
		&G^{-1}\left[\ddot{\tilde b}_1^{(1)} +iGe^{-i\pi\Phi/\Phi_0} \dot{\tilde
		b}_2^{(1)}   \right] = \\
		&-\left[i\tilde\omega- 2i\left| \tilde b_1^{(0)} \right|^2 + 
		\gamma \right]\dot{\tilde b}_1^{(0)} -i \left[1- \left( \tilde b_1^{(0)} 
		\right)^2  \right] 
		\left(\dot{\tilde b}_1^{(0)}\right)^*. \nonumber
	\end{align}
A second expression exist with $b_1\leftrightarrow b_2$ and $\Phi\to-\Phi$. This
can be used to eliminate $\dot{\tilde b}_2^{(1)}$ in
Eq.\eqref{eq:LCfirstorderG}. Inserting the expressions for the lowest order
terms and rewriting the products of harmonic functions we arrive at
	\begin{align}
		&\frac{1}{G}\left[\ddot{\tilde b}_1^{(1)} +G^2 \tilde b_1^{(1)} \right] = 
		-2\chi \dot{\tilde b}_1^{(0)} - \frac{|\beta|^2}{2}\beta
		e^{-i\pi\frac{\Phi}{\Phi_0}} G \cos[3Gt], \nonumber\\
		&\chi \equiv i\tilde\omega +\gamma - i\frac{3|\beta|^2}{4}- 
		i\frac{\beta^*}{2\beta}\left(e^{2i\pi\Phi/\Phi_0} -1 \right).
	\end{align}
There is a similar equation for $\tilde b_2$ with the term proportional to
$\cos[3Gt]$ replaced by $-i|\beta|^2\beta G\sin[3Gt]/2$. These equations
describe a driven harmonic oscillator. Since the term proportional to $\chi$
drives exactly at the resonance frequency, $G$, and the frequencies of the
higher order terms should only be multiples of $G$, we require $\chi=0$. This
sets $\beta$
	\begin{align}
	\begin{split}
		&|\beta_\pm|^2 = \frac{4}{3}\left[\pm \sqrt{\sin^2[\pi\Phi/\Phi_0] 
		-\gamma^2 } +\tilde\omega \right], \\
		&\gamma\tan[2\phi_\beta^\pm- \pi\Phi/\Phi_0] =
		\mp\sqrt{\sin^2[\pi\Phi/\Phi_0] 
		-\gamma^2 },
		\end{split}
	\end{align}
with $\phi_\beta^\pm$ the phase of $\beta_\pm$. We note that in this limit,
$|\beta|^2$ and therefore the leading order term of the average current is
independent of the coupling constant, $G$.
The first order terms are readily calculated
	\begin{align*}
		\tilde b_1^{(1)} = \frac{e^{-i\pi\frac{\Phi}{\Phi_0}}}{16G} 
		|\beta|^2\beta \cos[3Gt],\quad
		\tilde b_2^{(1)} = i\frac{|\beta|^2\beta}{16G} \sin[3Gt].
	\end{align*}
These variations have an extra factor of $i$ compared to the leading
order, and are therefore perpendicular to it in the complex plane. 

The correction to the number, $\delta \tilde n_j = \tilde n_j - \tilde
n_j^{(0)}$, is at least of the order $G^{-2}$. The phase between $\tilde b_1$
and $\tilde b_2$ is, up to first order, given by $\pi(2\Phi - \Phi_0)/2\Phi_0$.

\subsection{Numerics}
To validate the analytical results of the previous
subsection, we have performed a numerical analysis. We study the average current
through the two-mode LSID in the limit cycle regime (iv), and the trajectory of
the limit cycle.

The analysis is based on the numerical integration of the differential equations
in Eq.~\eqref{eq:doublecavityreduced}. The initial condition is chosen close to
$b_j=0$ and the parameters are chosen to achieve  the limit cycle regime. To
converge to
the limit cycle within a reasonable amount of integration time, we choose a
sufficiently large damping, $\gamma = 0.05$, which is still small enough for
all essential features to be as described in previous section. We integrate the
differential equation from $t=0$ up to $t=25/\gamma$. A time interval of $\delta
t = 1/\gamma$ at the end is used to represent the limit cycle, $b_j^{lc}(t)$. 

The data of $b_j^{lc}(t)$ is used to plot several quantities. We use the raw
data to demonstrate a few aspects of the limit cycle. The real and imaginary
parts of $b_j^{lc}(t)$, are plotted in a parametric plot to show its trajectory,
while the modulus and phase of $b_j^{lc}(t)$ are plotted as a function of time.
The frequency of the limit cycle is shifted from $G$ by
$G\delta\nu \simeq G^{-1}$.
We extract the value of $|\beta|^2$, by fitting $|b_j^{lc}(t)|^2$ to a function
of the form $|\beta_j|^2(1+ \sin[2G(1-\delta\nu)t +\kappa_j])$, corresponding to
the leading order solutions, $n_1^{(0)}$ and $n_2^{(0)}$. The higher orders are
small, being of order $G^{-2}$.

\begin{figure}
  \includegraphics{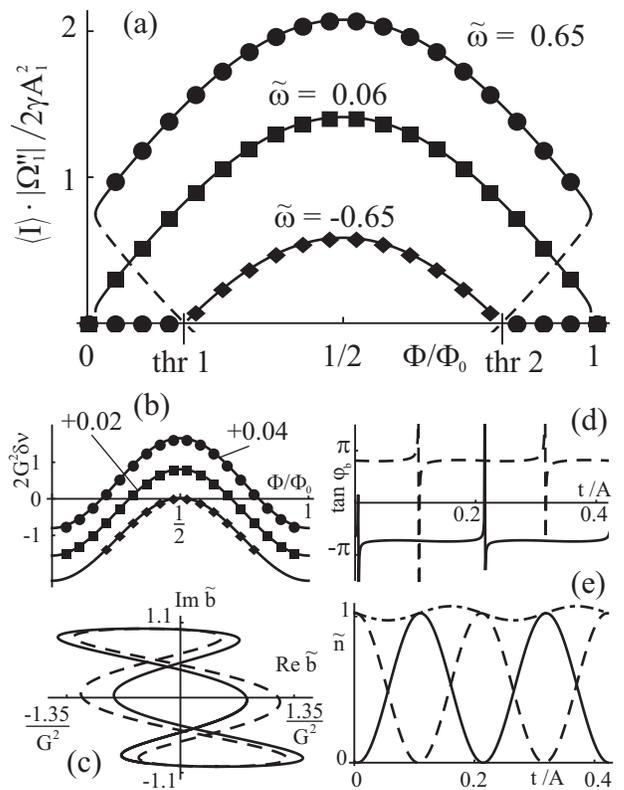}
  \caption{%
  Limit cycles in regime (iv) of the two-mode LSID, with $G=15$ and
$\gamma=0.05$. 
  (a) The average current as a function of flux. The circles, squares and
diamonds correspond to numerical results for three values of the detuning, as
indicated in the panel. The solid (dashed) lines are $|\beta_\pm|^2$,
corresponding to stable (unstable) limit cycle solutions. The values of the flux
labelled with `thr 1' and `thr 2' are lasing thresholds for the solutions with
$|\tilde\omega| = 0.56$. The solution at positive detuning is bistable in the
nonlasing regime, with a stable nonlasing state, ($I=0)$. This is similar to the
earlier discussed regime (iii). Close to $\Phi/\Phi_0= 0$ and $1$, a regime
similar to (i)b exists.
  (b) The variation of the relative cycle frequency shift $\delta\nu$ with flux.
The circles, squares and diamonds
correspond to the results in (a). The solid lines are fits with the function
$\eta_1\sin^2[\pi\Phi/\Phi_0]-\eta_2$. The coefficients ($\eta_1,\eta_2$) are
respectively given by $(0.074,0.064)$, $(0.071,0.067)$ and $(0.068,0.067)$. Two
curves are shifted by an amount indicated in the panel.
  (c) Trajectory of the limit cycle with $\Phi/\Phi_0=1/4$. The solid (dashed)
line corresponds to $\tilde b_1$ ($\tilde b_2$). The trajectory is rotated over
an angle of $0.87\pi$ ($1.12\pi$) to align the long axis of the cycle with the
vertical axis of the plot.
  (d) and (e) The optical phase and number of photons corresponding to the
trajectories in the limit cycle of (c). The dash-dot line in (e) is the sum of
the solid and dashed lines.
  }\label{fig:limitcycle}
\end{figure}

The results of the numerical analysis are shown in fig. ~\ref{fig:limitcycle}.
In panel (a), we show the average current through the two-mode LSID as a
function of the flux, $\Phi$, for three values of the detuning. We first remark
that the expression for $|\beta_+|^2$ nicely fits the numerical results.
Interestingly, we find two different regimes that remind of the regimes (ii) and
(iii) of the time independent states. Inevitably, this bistable regime also
involves an unstable limit cycle, which we expect to be represented by
$\beta_-$.
With this observation, we conclude that the limit cycle states display similar
parameter dependencies and properties as the time independent states 
investigated in Secs \ref{sec:singlemodedevice} and \ref{sec:dualmodeHJL}. In
particular, we find the regimes that are analogous to the regimes (i)b, (ii) and
(iii) and the
possibility of hysteresis as described in Sec.~\ref{sec:singlemodedevice}. With
this, we review our understanding of the regimes (ii)b, (iii)a (only at positive
$\omega$) and (iii)b, which were introduced in Sec.~\ref{sec:dualmodeHJL}C. We
only made notice of the existence of stationary states in these regimes, but we
expect that all these regimes also contain a stable and an unstable limit cycle
state, represented by $\beta_\pm$.

Panel (b) of fig.~\ref{fig:limitcycle} shows the relative shift of the cycle
frequency $\delta\nu = (1 -\omega_c/G)$, which is of
the order of $G^{-2}$. 

In the panels (c) - (e), the raw data is used to show the trajectory of the
limit cycle in a parametric plot, and the modulus and phase as a function of
time. The trajectory matches the prediction of the previous section. The long
axis of the paths correspond to the leading orders $b_j^{(0)}$, while the short
axis corresponds to the first order corrections, $b_j^{(1)}$. The difference in
shape between the trajectories of $\tilde b_1$ and $\tilde b_2$ result from the
corrections in the perturbation expansion of order $G^{-2}$ and higher. In the
panel (e), the moduli $|\tilde b_j(t)|^2$ are shown separately and as a sum,
$\tilde n_1(t)+\tilde n_2(t)$, which is proportional to the current. The
oscillation amplitude of the current depends on the relative phase of the
$b_j(t)$. 

We have described the limit cycles in the two-mode LSID. The dependence of the
limit cycle states on flux and detuning is rather similar to that of the time
independent stationary states. 
Generally, the emission spectrum in this case consists of a comb of equally
separated frequencies $\omega_n = eV/\hbar + \omega_c n$. 
Interestingly, in the limit of strong coupling the emission spectrum consists of
two discrete frequencies corresponding to $n=\pm 1$. This is therefore a dual
mode lasing state, in contrast to the states in regime (ii) that are single mode
lasing states. The two-mode LSID can thus lase at a single frequency, at
$\tilde\omega \simeq\pm G$, or at two frequencies at $\tilde\omega\lesssim 1$.
We stress that the occurrence of the dual mode lasing regime (iv), is crucially
related to the coupling of the superconductors to the resonator modes. Without
this coupling, we cannot use the flux to create the instability of the regime
(iv), that results in the dual mode lasing.

\section{Conclusions}\label{sec:conclusion}
We summarize the results of the article and sketch some prospectives of
HJL-based devices. 

We have studied two device setups reminiscent of a superconducting quantum
interference device (SQUID), where the regular Josephson junctions are replaced
by HJLs: the groups of quantum emitters, emitting in a resonator mode, of which
the optically active eigenstates are coupled to both superconducting leads. In
the first setup investigated, both groups of quantum emitters emit in a single
resonator mode, while in the second setup they emit in two separate resonator
modes, which are coupled optically. These setups were referred to as
respectively, `single mode LSID' and `two-mode LSID'. In both devices parameter
regimes exist that support lasing. The occurrence of nonlasing, lasing and
multistable regimes is equivalent to what is found in a regular HJL.
Additionally, the  LSIDs also depend on the magnetic flux that threads the
superconducting loop of the SQUID. It was found that the LSIDs can operate as a
flux tunable regular single mode HJL. Indeed, parameter regimes exist, where the
lasing in the LSIDs can be turned on and off by changing the magnetic flux
only. In this context, the occurence of bistable regimes leads for certain
parameter regimes to hysteretic behaviour upon performing flux sweeps.

The two-mode LSID has been studied in the weak and the strong coupling limit and
for a symmetric choice of parameters. In the weak coupling limit, the device is
equivalent to two single HJLs that perturb each other only slightly. A weak
dependence on the flux is found. In the strong coupling limit, the device
develops lasing instabilities at detunings of the order of the coupling
constant, both positive and negative. At these values of the detuning, the
device is similar to a single mode LSID. Studying the symmetric choice of
parameters revealed a new lasing instability in the vicinity of zero detuning,
which was investigated in the limit of strong coupling. Here, the device
exhibits lasing that is predominantly occuring at two frequencies, which are
separated by approximately twice the coupling strength. For such dual mode
lasing, there are regimes similar to the ones of the time independent states:  a
nonlasing, a lasing and a bistable one. 

The connection between superconductivity and optics achieved with the HJL
devices promises a set of novel applications, this article providing an example
thereof. With these prospects, the emerging field of superconducting
opto-electronics looks rather promising.

Even more possibilities would emerge for arrays of HJLs.  It is easy to extend
the design idea of the two-mode LSID to an $n$-mode LSID.

The setup for such an $n$-mode LSID consists of $n$ HJLs in parallel, all
sharing the same pair of superconducting electrodes. This guarantees that the
devices are driven at the same frequency. 
Note that there are $n-1$ superconducting loops in this circuit, making it
possible to tune the superconducting phase differences of each HJL.
An optical coupling between the nearest HJLs is provided. The dynamics 
is described by a set of $2n$ equations, those generalize 
Eq.~\eqref{eq:doublecavityreduced}. Each of these equations  contains two
coupling terms. Linearized equations give $n$ resonant modes. If the detuning
matches the resonant frequencies, we expect a single-mode lasing. Otherwise, the
lasing regimes may become complex, involving limit cycles and perhaps even
chaos. The lasing regimes can be tuned with changing the fluxes in the loops.

The $n$-mode LSID is a fairly straigthforward extension of the ideas of this
article. It is reminiscent to the arrays of Josephson
junctions\cite{fazio} that can be regarded as a realization of artificial
solids. Similarly to the Josephson junction arrays, there are rich design
possibilities for such HJL devices. One could design any kind of setup with
superconducting loops, in 1D, 2D or even 3D, incorporate as many HJLs as
necessary, and couple those optically with each other. The coupling does not
even have to be limited to the nearest neighbours. In principle, it can be
realized with any number of
neighbours, and with varying coupling strengths. This would open up a new field
of research, where the physical phenomena typical for in Josephson
arrays\cite{qpt,vortices,bkt} merge with optics and lasing. 

\acknowledgments
We acknowledge financial support from the Dutch Science Foundation NWO/FOM.

\end{document}